\documentclass[prl,twocolumn,showpacs,superscriptaddress,preprintnumbers,amssymb]{revtex4}
\usepackage{graphicx}% Include figure files

\usepackage{dcolumn}% Align table columns on decimal point
\usepackage{bm}% bold math

\newcommand{\beq}{\begin{equation}}
\newcommand{\eeq}{\end{equation}}
\newcommand{\beqn}{\begin{eqnarray}}
\newcommand{\eeqn}{\end{eqnarray}}

\begin{document}

%\title{Nonperturbative effects of Topological $\Theta$-term}

\title{Nonperturbative effects of a Topological $\Theta$-term
on Principal Chiral Nonlinear Sigma Models in (2+1) dimensions}

%\author{Authors}

\author{Cenke Xu}

\author{Andreas W. W. Ludwig}
\affiliation{Department of Physics, University of California,
Santa Barbara, CA 93106}

\begin{abstract}

We study the effects of a topological $\Theta$-term on 2+1
dimensional principal chiral models (PCM), which are nonlinear
sigma models defined on Lie group manifolds. We find that when
$\Theta = \pi$, the nature of the disordered phase of the
principal chiral model is strongly affected by the topological
term: it is either a gapless conformal field theory, or it is
gapped and two-fold degenerate. The result of our paper can be
used to analyze the boundary states of three dimensional symmetry
protected topological phases.

\end{abstract}

\date{\today}

\maketitle

\noindent \underbar{\it Introduction:} It is well known that
topological terms in field theories are responsible for many
profound phenomena in condensed matter physics. For instance, the
one-dimensional SU(2) spin-1/2 Heisenberg quantum spin chain is
known to be described by the (1+1)-dimensional  O(3) Nonlinear
Sigma Model [(1+1)d NLSM] with a $\Theta$-term, for a (real)
three-component unit vector ${\vec n}$ on the two-dimensional
sphere $S^2$ \cite{affleck,haldane1,haldane2} with action \beqn S
= \int d^2x \ \frac{1}{g} (\partial_\mu \vec{n})^2 +
\frac{i\Theta}{8 \pi} \epsilon_{\mu\nu} \epsilon_{abc} \ n^a
\partial_\mu n^b \partial_\nu n^c.
\label{1do3} \eeqn Throughout this article we will always work in
{\it imaginary time}. The $\Theta$-term contributes a factor
$\exp(i\Theta q)$ to the partition function of the NLSM for every
field configuration $\vec{n}(x)$ in (1+1)d space-time which has
instantons of `topological charge' $q$. For a spin-$s$ chain
$\Theta = 2\pi s$, which leads to the qualitative difference
between integer and half-integer spin chains, due to the
constructive and destructive interference between even and odd
number of instantons for  the two different values of $\Theta$.

A similar NLSM with a $\Theta$-term can be
used\cite{khmelnitskii,pruisken1,pruisken2} to describe the
integer quantum Hall state in two dimensions, where the target
space is $\mathrm{U(2N)/[U(N) \times U(N)]}$ and\cite{ReplicaSUSY}
$N\to 0$. (When $N = 1$ this model reduces to the O(3) NLSM in
Eq.~\ref{1do3}.) The quantized Hall
conductivity\cite{UnitsOfSigmaxy} $\sigma_{xy}$  is precisely
identified with $\Theta = 2\pi k$ where $k$ is integer. The
plateau transition between different integer quantum Hall states
is described by the same NLSM, but with $\Theta = \pi (2k+1)$.

In the present article, we will study the (2+1)-d Principal Chiral
Nonlinear Sigma Model (PCM) with a theta term, which has the
action \beqn S &=& \int d\tau d^2x \ \frac{1}{g}
\mathrm{tr}[\partial_\mu U^\dagger
\partial_\mu U] \cr\cr &+& \frac{i \Theta}{24 \pi^2}
\epsilon_{\mu \nu \rho}  \mathrm{tr}[(U^\dagger \partial_{\mu} U)(
U^\dagger \partial_\nu U) (U^\dagger \partial_\rho U)]. \label{pc}
\eeqn where $U$ is a group element that belongs to a (simple)
compact Lie group $G$, such as SU(N), SO(N), Sp(N). All these
groups have nontrivial homotopy group $\pi_3[G] = Z$, which
implies that the corresponding PCMs possess instantons in (2+1)-d
space-time, and that a $\Theta$-term can be added to the action
(as in Eq. \ref{pc}). For arbitrary values of $\Theta$ the PCM in
Eq.~\ref{pc} is invariant under a $G_L \times G_R$ symmetry,
denoting left and right multiplication of $U$ by group elements.
When $\Theta = \pi k$ with integer $k$, the system also has other
discrete symmetries such as reflection $x\rightarrow -x$, or
time-reversal that transforms $i \rightarrow - i$ (we assume $U$
carries a trivial representation of time-reversal). Thus these
discrete symmetries guarantee that when $\Theta = \pi k$, $\Theta$
does not flow under the renormalization group (RG), while for any
other value $\Theta$ is not forbidden to flow.

Below, we will present our results first for the special case of
$G = \mathrm{SU(2)}$. Subsequently, we will explain that our
arguments and conclusions are in fact generally applicable to any
(simple) compact Lie group $G$. In the case of $G=SU(2)$ one can
parametrize any group element in terms of a four-dimensional
(real) unit vector ${\vec \phi}^t =(\phi^0, \phi^1, \phi^2,
\phi^3)$ on the 3-dimensional sphere $S^3$ as $U = \phi^0 + i
\phi^1 \sigma^x +  i \phi^2 \sigma^y + i \phi^3 \sigma^z$. Now,
the $G_L \times G_R=$ $SU(2)_L\times SU(2)_R$ symmetry group is
isomorphic to $SO(4)$. This implies that for $G=SU(2)$, the PCM in
Eq.~\ref{pc} is equivalent to the (2+1)-d O(4) NLSM with action
\beqn S = \int d^3x \ \frac{1}{g} (\partial_\mu \vec{\phi})^2 +
\frac{i\Theta}{12\pi^2} \epsilon_{\mu\nu\rho} \epsilon_{abcd} \
\phi^a \partial_\mu \phi^b
\partial_\nu \phi^c \partial_\rho \phi^d.
\label{theta} \eeqn

The goal of this paper is to provide a non-perturbative argument
for the phase diagram of the (2+1)-d PCM with $\Theta-$term,
Eq.~\ref{pc}, in terms of the two coupling constants $g$ and
$\Theta$. Our result for this phase diagram is depicted in
Fig.~\ref{rg} below. Note that in the absence of space-time
boundaries we are allowed to compactify (2+1)-d space-time into a
three dimensional sphere $S^3$. First consider the case where
$\Theta$ is an integer multiple of $2\pi$, $\Theta = 2 \pi k$, so
that the $\Theta$-term contributes a factor of unity to the
partition function for any nontrivial instanton configuration in
the space-time. Thus, in the absence of boundaries in space-time
the phase diagram of PCM in Eq.~\ref{theta} at $\Theta = 2 \pi k$
is identical to the model at $\Theta=0$. For small values of the
coupling $g$, the system is in an O(4) ordered phase with nonzero
order parameter $\langle \vec{\phi} \rangle \neq 0$ and three
gapless Goldstone modes. For large $g$, on the other hand, the
system is in a quantum disordered phase with a non-degenerate
ground state and a fully gapped spectrum. The quantum phase
transition between the O(4) ordered and disordered phases is an
ordinary 2nd order transition in the 3D O(4) Wilson-Fisher (WF)
universality class.

The presence of a $\Theta$-term is not expected to affect the
ordered phase, because in it instantons are suppressed. Therefore,
it can only play a role for the transition into the disordered
phase and in the disordered phase itself. In order to understand
the disordered phase and the phase transition in the presence of a
$\Theta-$term, standard perturbative methods fail.
%fail\cite{FootnotePertFail}.
Thus in order to understand the disordered phase of the PCM in
Eq.~\ref{theta}, a non-perturbative argument must be developed,
which is what we do in this article. Our conclusion is that there
are two possibilities for the disordered phase of the PCM with a
$\Theta-$term when $\Theta = \pi$ (Eq.~\ref{theta} and
Eq.~\ref{pc}): It is either a gapless phase with power-law
correlations for the fields $U$ (or $\vec{\phi}$), or it is a
gapped phase but with a two-fold ground state degeneracy.

Very recently the new concept of symmetry protected topological
(SPT) phases appeared, which has attracted significant
attention~\cite{wentopo01,wentopo02,levinstern,liuwen,levingu,luashvin,levinsenthil,ashvinsenthil,xu3dspt,xu2003b,xusenthil}.
The PCM in Eq.~\ref{pc} has been used as a general formalism to
describe SPT
phases~\cite{wentopo01,liuwen,levinsenthil,xu3dspt,xu2003b,xusenthil}.
Our conclusion is especially useful for studying the edge states
of 3+1d symmetry protected topological phases, which in many cases
are precisely described by Eq.~\ref{pc} at $\Theta =
\pi$~\cite{ashvinsenthil,xu3dspt}.

\vskip .1cm \noindent
%\underbar{\it (1+1)-d $O(3)$ NLSM at $\Theta=\pi$:}
\underbar{\it $O(3)$ NLSM at $\Theta=\pi$ in (1+1)  dimensions:}
Before we start our argument in (2+1)-d, let us first consider the
(1+1)-d O(3) NLSM with the $\Theta$-term in Eq.~\ref{1do3}, and
focus on $\Theta = \pi$. Since this model describes the SU(2)
spin-1/2 chain, we know from the Lieb-Schultz-Mattis (LSM) theorem
\cite{LSM} that this NLSM  is either a gapless conformal field
theory (CFT), or gapped but two-fold degenerate. Since the goal of
this paper is to understand the (2+1)-d PCM field theories of
Eq.~\ref{pc} and Eq.~\ref{theta} without recourse to any lattice
spin model representations, we will first use a new argument to
understand the behavior of the (1+1)-d O(3) NLSM without using any
lattice representations such as spin chains. Subsequently, we will
generalize this argument to the (2+1)-d models.

\noindent Our argument proceeds in four steps:

{\it Step (1).} In order to understand the O(3) NLSM
Eq.~\ref{1do3} at $\Theta = \pi$, let us first look at $\Theta =
0$ and $\Theta = 2\pi$. At these values of $\Theta$, the system
also has the discrete symmetry $\vec{n} \rightarrow -\vec{n}$, in
addition to the SO(3) rotation symmetry. The bulk spectra for
$\Theta = 0$ and $\Theta = 2\pi$ are identical, possessing a
non-degenerate ground state and a gap to all excitations.

{\it Step (2).} Now let us consider the system on a spatial
interval with open boundaries at $x = 0$ and $x = L$. Although the
models with $\Theta = 0$ and $\Theta = 2\pi$ have identical bulk
spectra, they behave very differently at the boundaries. Since the
bulk is gapped, we can safely ignore the bulk, and focus on the
boundary because the gap in the bulk will protect the effective
boundary theory from any singular contributions. Since the
boundary is a point in space, it is effectively described by a
(0+1)-d O(3) NLSM model. When $\Theta = 0$ this (0+1)-d NLSM model
is completely trivial. However, when $\Theta = 2\pi$, the
$\Theta$-term in Eq.~\ref{1do3} can be viewed as the O(3)
Wess-Zumino-Witten (WZW) term for the (0+1)-d O(3) NLSM model at
each of the two boundaries, at $x = 0$ and at $L$: \beqn &&
\int_0^L dx \int d\tau \frac{i 2\pi}{8 \pi} \epsilon_{\mu\nu}
\epsilon_{abc} \ n^a
\partial_\mu n^b \partial_\nu n^c \cr\cr && = \mathrm{WZW}_{0} -
\mathrm{WZW}_{L}. \label{wzw} \eeqn The WZW term for a 0+1
dimensional O(3) NLSM, appearing on the right hand side, is
defined as follows: In the (0+1)-d NLSM the O(3) vector $\vec{n}$
is a function only of imaginary time $\tau$. Consider a periodic
evolution of $\vec{n}(\tau)$, namely $\vec{n}_{\tau = 0} =
\vec{n}_{\tau = \beta}$. Then $\vec{n}$ is a mapping from a closed
loop $S^1$ parametrized by $\tau \in [0, \beta]$ to the target
space $S^2$. The WZW term is defined as the solid angle on the
target space $S^2$ enclosed by the closed loop $\tau \in [0,
\beta]$. The WZW term at level $k$ can be explicitly written as
\beqn \mathrm{WZW} &=& 2\pi \int d\tau \int_0^1 du  \
\frac{ik}{8\pi} \epsilon_{abc}\epsilon_{\mu\nu} n^a
\partial_\mu n^b \partial_\nu n^c. \label{wzw2} \eeqn Here, the
function $\vec{n}(\tau)$ has been extended to a mapping
$\vec{n}(\tau, u)$ from a disc $(\tau, u)$, where $0 \leq u \leq
1$ and $\tau \in S^1$, to the target $S^2$. This extension has one
constraint: $\vec{n}(\tau, 1) = \vec{n}(\tau)$, $\vec{n}(\tau, 0)
= \hat{z}$. Unlike the $\Theta$-term, the coefficient $k$ in
Eq.~\ref{wzw2} has to be an integer\cite{FootnoteWZWlevelInteger},
regardless of whether any discrete symmetry is present or not. By
simply identifying $u$ with $x$, one arrives at Eq.~\ref{wzw}.

It is well known that if a (0+1)-d O(3) NLSM, describing the
quantum mechanics of a point particle on a sphere, has a WZW term
at level $k$, the ground state of this quantum mechanics is
$(k+1)$-fold degenerate. In fact, the ground state of the (0+1)-d
O(3) NLSM with a WZW term at level $k$ precisely describes a
single SU(2) spin with $S = k/2$. In Eq.~\ref{wzw}, the WZW term
at each boundary is at level $k = 1$. This implies that the model
in Eq.~\ref{1do3} has two fold degeneracy at each boundary when
$\Theta = 2\pi$. This conclusion again agrees with Haldane's
conjecture, which states that the model with $\Theta = 2\pi$
describes the spin-1 chain. Moreover it recovers the well-known
fact that the Haldane phase of the spin-1 chain has an unpaired
spin-1/2 degree of freedom at each of its boundaries
\cite{hagiwara1990,glarum1991,ng1994}.

{\it Step (3).} Now let us tune $\Theta$ in Eq.~\ref{1do3}
continuously from $2\pi$ to $0$. Then the spin-1/2 boundary state
has to disappear at a certain value of $\Theta$. When $\Theta$ is
tuned away from $2\pi$, the discrete symmetry $\vec{n} \rightarrow
- \vec{n}$ of the system is broken. One important fact is that the
spin-1/2 boundary state cannot be destroyed without going through
a bulk transition, even when the discrete symmetry is broken. This
is because, given a single spin-1/2, as long as the SO(3) symmetry
is preserved, the spin-1/2 doublet degeneracy cannot be lifted.
This conclusion can also be drawn by noticing that the coefficient
of the WZW term has to be quantized, no matter whether the
discrete symmetry is broken or not.

$\Theta = k \pi$ with integer $k$ is a fixed point under
renormalization group, while with any other value $\Theta$ in
principle can flow under RG. We should use the fixed point values
of $\Theta$ to derive the edge WZW model. The paragraph above
implies that when $\Theta = 2\pi \pm \epsilon$, the edge state is
identical with the edge state with $\Theta = 2\pi$. If one
introduces the standard CP$^1$ representation like in
Ref.~\cite{ng1994}, the $\Theta-$term in Eq.~\ref{1do3} becomes
the $\Theta-$term of the U(1) gauge field, and when $\Theta \in
(-\pi, 3\pi)$ there is always a gauge charge-1 (spin-1/2)
localized at the boundary~\cite{coleman}.

{\it Step (4).} We have concluded that in order to destroy the
boundary spin-1/2 state, a bulk transition has to occur. Here we
assume the simplest case, $i.e.$ that there is one single
transition between $\Theta = 0$ and $2\pi$. Then in this case
there are exactly two possibilities for this bulk transition:

{\it 4A.} The transition is of second order, meaning that the bulk
gap closes continuously for some values of $\Theta$ between
$\Theta = 0$ and $2\pi$. Because the bulk spectrum is identical
for $\Theta$ and $(2\pi - \Theta)$, this transition has to occur
at $\Theta = \pi$ if the bulk gap closes only at {\it one} value
of $\Theta$. This implies that the bulk is gapless when $\Theta =
\pi$. When $\Theta$ is approaching $\pi$ from $2\pi$, the boundary
spin-1/2 state will become more and more delocalized, and
eventually gets absorbed by the gapless bulk states at $\Theta =
\pi$.

\begin{figure}
\begin{center}
\includegraphics[width=3.2 in]{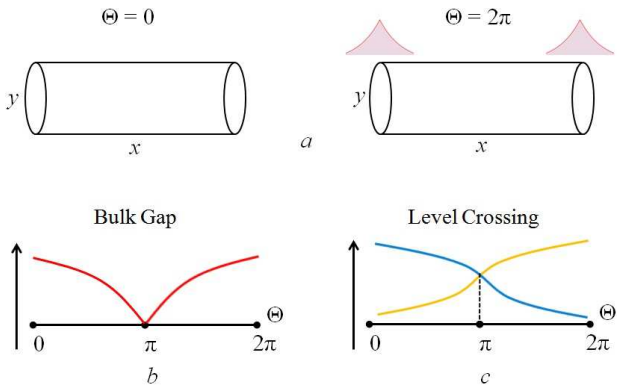}
\caption{$(a)$. We compactify the space of model Eq.~\ref{pc} and
Eq.~\ref{theta} to a two dimensional cylinder. When $\Theta =
2\pi$ there are gapless boundary states localized at the two
boundaries. $(b)$. The first possibility when we tune $\Theta$
from $2\pi$ to $0$, the bulk gap closes at $\Theta = \pi$. $(c)$.
The second possibility, the two states at $\Theta = 0$ and $\Theta
= 2\pi$ have level crossing at $\Theta = \pi$.} \label{tuning}
\end{center}
\end{figure}

{\it 4B.} The transition is of first order, meaning there is
always a bulk gap. However, at this first order transition, the
two phases with $\Theta = 0$ and $\Theta = 2\pi$ will have a
crossing of their ground state energies, and this level crossing
also has to occur at $\Theta = \pi$. This implies that the ground
state at $\Theta = \pi$ is two-fold degenerate. In this case, when
$\Theta$ is tuned from $2\pi$ to $\pi$, the boundary spin-1/2
states will never delocalize, they will simply disappear abruptly
at $\Theta = \pi$. An example of this phenomenon is the (1+1)-d
CP$^N$ model which is known\cite{witten,coleman,Affleck1988} to
have a first order transition at $\Theta = \pi$ when $N\geq 3$.

These two possibilities that we have arrived at above are
completely consistent with the conclusion drawn from the LSM
theorem for the SU(2) spin-1/2 chain.

\vskip .1cm \noindent
%\underbar{\it (2+1)-d $O(4)$ NLSM at $\Theta=\pi$:}
\underbar{\it $O(4)$ NLSM at $\Theta=\pi$ in (2+1) dimensions:} We
will now generalize the arguments given in the previous paragraph
to the (2+1)-d O(4) NLSM with a $\Theta$-term, Eq.~\ref{theta}.
Since we are only interested in the nature of the disordered phase
of Eq.~\ref{theta}, we will consider the case with large values of
the coupling $g$.

{\it Step (1).} In order to investigate the disordered phase at
$\Theta = \pi$, we first look, as before, at $\Theta = 0$ and
$2\pi$. Again, in the bulk these two disordered phases are both
gapped with a non-degenerate ground state, while they have
different boundary states. In order to look at the boundary
states, we let the $x$ direction be a finite interval $0 \leq x
\leq L$ while the $y$ direction is periodic, so that the system is
defined on a finite 2d cylinder (Fig.~\ref{tuning}$a$).

{\it Step (2).} At each boundary located at $x = 0$ and $x = L$
there is a (1+1)-d theory defined on $(y, \tau)$-space-time. Since
the bulk is gapped when $\Theta = 0$ and $2\pi$, the kinetic term
of the effective boundary theory is still that of a local O(4)
NLSM, Eq.~\ref{theta}, but now in (1+1)-d. When $\Theta = 0$,
there is no nontrivial topological term at the boundary. However,
when $\Theta = 2\pi$, the $\Theta$-term of the (2+1)-d bulk $O(4)$
NLSM, Eq.~\ref{theta}, can as before be viewed as a WZW term of
the (1+1)-d $O(4)$ NLSM appearing on the boundary. Thus, the
boundary theory is described by the following (1+1)-d O(4) NLSM
with a WZW term: \beqn && S = \int dy d\tau \ \frac{1}{g}
(\partial_\mu \vec{\phi})^2 \cr\cr
  &+& \frac{i 2\pi k}{12\pi^2}
\int dy d\tau \int_0^1 du  \epsilon_{\mu\nu\rho} \epsilon_{abcd} \
\phi^a
\partial_\mu \phi^b \partial_\nu \phi^c \partial_\rho \phi^d.
\label{1do4} \eeqn When $\Theta = 2\pi$, the WZW term in
Eq.~\ref{1do4} has level $k = 1$. It is well
known\cite{Witten1984,KnizhnikZamolodchikov1984} that the
long-distance behavior of this (1+1)-d O(4) NLSM with level $k=1$
WZW term is controlled by a stable fixed point at finite $g^\ast$,
and that this fixed point is precisely the SU(2)$_1$ CFT which
describes the nearest neighbor spin-1/2 Heisenberg chain. When
$\Theta = 2\pi k$ and $k=$ integer, the boundary is described at
long scales by the $SU(2)_k$
CFT\cite{Witten1984,KnizhnikZamolodchikov1984}. Thus once again,
when $\Theta$ is a nonzero integer multiple of $2 \pi$ the system
possesses nontrivial gapless boundary states.

{\it Step (3).} The same strategy that we used before can now be
applied: When we tune $\Theta$ continuously from $2\pi$ to 0, then
the boundary state has to disappear through a bulk phase
transition. This is because SO(4) symmetry of the (1+1)-d theory
in Eq.~\ref{1do4} is, as mentioned in the introduction, isomorphic
to the $SU(2)_L\times SU(2)_R$ symmetry of the PCM. It is this
symmetry that protects the finite-coupling fixed point at $g=g^*$
of the boundary NLSM, Eq.\ref{1do4}, from being gapped out. In
order to gap out this fixed point CFT, we need to break the
$SU(2)_L\times SU(2)_R$ symmetry down to the diagonal SU(2)
symmetry, $i.e.$ we need to induce a relevant back-scattering
between left and right moving boundary modes. However, since our
model Eq.~\ref{theta} has $O(4)\sim SU(2)_L\times SU(2)_R$
symmetry for any value of the coupling $g$, such backscattering
processes are absent. Thus, although tuning $\Theta$ away from
$2\pi$ breaks a discrete symmetry of the system, the boundary CFT
cannot be gapped out without going through a bulk transition.

{\it Step (4).} Since the boundary states can only be destroyed
through a bulk transition, there are the following two
possibilities for this transition:

{\it 4A.} This bulk transition is of second order, and it has to
occur at $\Theta = \pi$. This implies that the disordered phase of
the PCM in  Eq.~\ref{theta} is gapless at $\Theta = \pi$. Since
the transition is of second order, the second derivative
$\partial^2 E(\Theta)/\partial \Theta^2$ of the ground state
energy $E(\Theta)$ has a singularity at $\Theta = \pi$
(Fig.~\ref{tuning}$b$).

{\it 4B.} This bulk transition is first order and occurs at
$\Theta = \pi$. At this transition the two gapped phases with
$\Theta = 0$ and $\Theta = 2\pi$ will have a crossing of their
ground state energies, and this level crossing has to appear at
$\Theta=\pi$. This implies, as before, a gapped spectrum and a
two-fold degenerate ground state at $\Theta = \pi$. In this case,
the ground state energy $E(\Theta)$ has a kink at $\Theta = \pi$,
$i.e.$ the first order derivative $\partial E(\Theta)/\partial
\Theta$ is discontinuous at $\Theta = \pi$ (Fig.~\ref{tuning}$c$).

\begin{figure}
\begin{center}
\includegraphics[width=3.3 in]{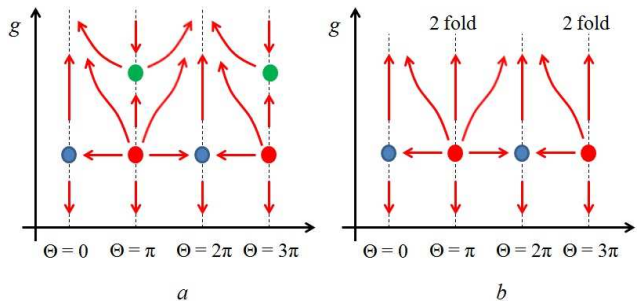}
\caption{The two possible RG flows for the coupling constants $g$
and $\Theta$ of the (2+1)-d PCM on a compact Lie group $G$ with
Theta term,  Eq.~\ref{pc} (and Eq.~\ref{theta} for $G=SU(2)$).
There is always an ordered phase with small $g$. For $G=SU(2)$,
the phase transition is in the conventional three-dimensional O(4)
Wilson-Fisher universality class when $\Theta = 0$ and $2\pi$,
while the fixed points at $\Theta=\pi$ are presumably in the
different universality classes. } \label{rg}
\end{center}
\end{figure}

\vskip .1cm \noindent
%\underbar{\it (2+1)-d $O(4)$ NLSM at $\Theta=\pi$:}
\underbar{\it PCM on group $G$ at $\Theta=\pi$ in (2+1)
dimensions:} It is straightforward to generalize these arguments
to other (2+1)-d PCMs with a Theta term, as in Eq.~\ref{pc},
defined on more general compact Lie group manifolds such as e.g.
$G= SU(N), SO(N)$ and $Sp(N)$. The key argument rests on the
gaplessness of the CFT that describes the long-distance behavior
of the (1+1)-d PCM with WZW term at level $k$, which appears at
the boundary of the (2+1)-d bulk PCM at $\Theta= 2 \pi k$. The
gaplessness of this CFT is protected by the $G_L\times G_R$
symmetry of the PCM with WZW term, which
forbids\cite{KnizhnikZamolodchikov1984} all operators that are
relevant in the RG sense from appearing in the action.
%which forbids\cite{KnizhnikZamolodchikov1984} all operators
%which are relevant in the renormalization group (RG) sense.

Based on these arguments we obtain the two possibilities for the
RG flow diagram of the two coupling constants $g$ and $\Theta$ of
model Eq.~\ref{pc} sketched in Fig.~\ref{rg}.

{\it Another argument}

Now let us use a different argument to achieve the same
conclusion. The goal is still to prove the bulk spectrum of
Eq.~\ref{theta} is either gapless or degenerate at $\Theta = \pi$.
To make this conclusion, we only need to argue that {\it if
Eq.~\ref{theta} is gapped at $\Theta = \pi$, it must be
degenerate}. Thus let us assume it is gapped, then after rescaling
the system to the infrared limit this gap becomes infinity, which
means that $g$ cannot flow to any finite fixed point, nor can it
flow to zero, because in either case the system will be gapless.
Thus the coupling $g$ can be taken to infinity effectively. Then
we can ignore the first term in Eq.~\ref{theta}, and focus on the
second, topological term only. Since we are interested in the bulk
properties, we may compactify two dimensional space to a
two-sphere $S^2$; $i.e.$, any spatial configuration of
$\vec{\phi}(x,y,\tau)$ {\it at fixed time} is a mapping from the
two-sphere $S^2$ to the target space $S^3$. Now, we can write the
$\Theta$-term in the following form: \beqn && \int d\tau \ \left (
\Theta\mathrm{-term} \right ) = \int d\tau \ \left ( i\Theta
\partial_\tau \Phi\right ) , \cr\cr \Phi(\tau) &=& \int d^2x \int
du \ \frac{1}{12 \pi^2} \ \epsilon_{abcd}\epsilon_{\mu\nu\rho}
\phi^a \partial_\mu \phi^b
\partial_\nu \phi^c \partial_\rho \phi^d. \label{Phi}
\eeqn
%$\Phi$ is an analogue of the WZW term in Eq.~\ref{wzw}.
%In Eq.~\ref{Phi} we have extended $\vec{\phi}(x,y)$ to a mapping from
%the three dimensional ball $D^3$ parameterized by $(x, y, u)$ to
%the target space $S^3$, and the real space $S^2$ is the boundary
%of $D^3$.
%Just like the WZW term, the only constraint on this
%extended mapping is $\vec{\phi}(x, y, 1) = \vec{\phi}(x, y)$, $
%\vec{\phi}(x, y, 0) = (0 , 0 , 0, 1)$.
Here, $\Phi$ plays the role of the WZW term in Eq.~\ref{wzw}. In
Eq.~\ref{Phi} we have extended the spatial dependence of
$\vec{\phi}(x,y,\tau)$ at fixed time to a mapping from the three
dimensional ball $D^3$ parameterized by $(x, y, u)$ to the target
space $S^3$, so that the spatial two-sphere $S^2$ is the boundary
of $D^3$. Just like the WZW term, the only constraint on this
extended mapping is $\vec{\phi}(x, y, 1,\tau) = \vec{\phi}(x,
y,\tau)$, and $ \vec{\phi}(x, y, 0,\tau) = (0 , 0 , 0, 1)$, for
all times $\tau$.

\begin{figure}
\begin{center}
\includegraphics[width=3.0 in]{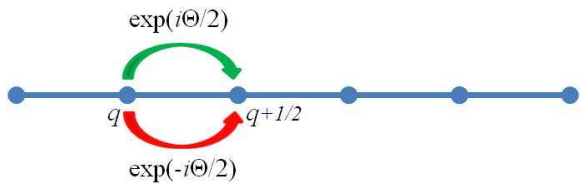}
\caption{When the O(4) symmetry of Eq.~\ref{theta} is broken down
to O(3)$\times Z_2$, the Skyrmion number is quantized on space
$S^2$. We map Eq.~\ref{Philag} to a one dimensional tight-binding
model. Hopping between nearest neighbor sites corresponds to
changing the Skyrmion number by 1. The $\Theta$-term grants two
types of monopoles a factor $\exp(i\Theta/2)$ and $\exp(-
i\Theta/2)$ respectively, which forbids nearest neighbor hopping
when $\Theta = \pi$ due to destructive interference between these
two types of monopoles.} \label{monolattice}
\end{center}
\end{figure}

%Now $g$ is taken to infinity, the action Eq.~\ref{theta} is
%simplified to a quantum mechanics problem:
What is the physical meaning of $\Phi$? To see this most
intuitively, we can weakly break the O(4) symmetry down to the
$\mathrm{O(3)} \times Z_2$ symmetry like in
Ref.~\cite{senthilfisher}, and take $\vec{\phi} = (\phi^0,
\vec{n})$. Then $\Phi(\tau)$ precisely reduces to \beqn \Phi(\tau)
= \frac{1}{16\pi} \int dx dy \ \epsilon_{abc}\epsilon_{\mu\nu} n^a
\partial_\mu n^b \partial_\nu n^c,
\eeqn $i.e.$ $\Phi(\tau)$ is related to the Skyrmion number of
$\vec{n}$ evaluated {\it at fixed time $\tau$}. Once the O(4)
symmetry is broken down to O(3)$\times Z_2$, $\Phi(\tau)$ is no
longer a continuous variable, instead it is a discrete number,
because the O(3) Skyrmion number is quantized on the space $S^2$.
More precisely, when the instantaneous field $\vec{n}(x,y,\tau)$
at time $\tau$ has $q$ Skyrmions in 2d space, then $\Phi(\tau) = q
/2$. Thus this O(4) to O(3)$\times Z_2$ symmetry breaking is
equivalent to turning on a strong periodic potential $V(\Phi)$ in
Eq.~\ref{Phi} with periodicity $1/2$. Now, as $g$ is taken to
infinity, the action Eq.~\ref{theta} simplifies to that of a
single particle quantum mechanics problem: \beqn S = \int d\tau \
\frac{1}{2}m (\partial_\tau \Phi)^2 \pm i \Theta
\partial_\tau \Phi + V(\Phi), \ \ m \rightarrow 0,
\label{Philag} \eeqn where $V(\Phi)$ is the effective periodic
potential for $\Phi$, and the Lagrangian Eq.~\ref{Phi} reduces to
a one dimensional tight-binding model with lattice constant 1/2
(Fig.~\ref{monolattice}) $i.e.$ the size of the Brillouin zone of
this tight binding model is $4\pi$. Nearest neighbor hopping on
this lattice is physically equivalent to changing the Skyrmion
number on space $S^2$ by 1.

To change the Skyrmion number, one needs a space-time
hedgehog-like monopole configuration of $\vec{n}$. The core of the
monopole has $\phi^0 \neq 0$, and there are two different types of
monopole, depending on the sign of $\phi^0$ at the monopole core.
The $\Theta$-term will contribute a factor $\exp(i\Theta/2)$ and
$\exp(- i\Theta/2)$ to the monopole with $\phi^0 > 0$ and $\phi^0
< 0$ respectively (The $\pm$ sign in E.q~\ref{Philag}). Thus when
$\Theta = \pi$ these two types of monopoles have complete
destructive interference with each other, and it is forbidden to
change the Skyrmion number by 1, $i.e.$ hopping by one lattice
constant in Fig.~\ref{monolattice} is forbidden. However, hopping
by two lattice constants is still allowed, but the band structure
will be doubly degenerate, namely on this one dimensional lattice
the states with momentum $p = 0$ and $p = 2\pi$ are degenerate.
The wavefunctions of these two states are \beqn |0\rangle \sim
\sum_{q} | q \rangle, \ \ \ \ |1\rangle \sim \sum_{q} (-1)^q | q
\rangle, \label{2states}\eeqn where $q$ is the Skyrmion number on
$S^2$.

Ref.~\cite{senthilfisher} has considered the O(4) to O(3)$\times
Z_2$ symmetry breaking in Eq.~\ref{theta}, and using a different
formalism the authors argued that the disordered phase has two
fold degeneracy at $\Theta = \pi$. Although the argument was
different, the two degenerate ground states derived in
Ref~\cite{senthilfisher} are precisely the two states in
Eq.~\ref{2states}.

In summary, using two different nonperturbative arguments, we
conclude that the topological $\Theta-$term can drastically change
the dynamics of the quantum disordered phases of PCMs in (2+1)-d
space-time, when $\Theta = \pi$. As we mentioned at the beginning
of this paper, the notion of a SPT phase has become
%a very
an important
%importance
concept in condensed matter theory. The key of constructing a SPT
phase  is to prove that  its
%boundary is either gapless or degenerate.
boundaries are either gapless or possess a degeneracy. The
boundary states of many (3+1)-d SPT phases can be mapped exactly
to a (2+1)-d PCMs with $\Theta =
\pi$~\cite{ashvinsenthil,xu3dspt}. Thus the result in the current
paper immediately concludes that the boundaries  of these SPT
phases are nontrivial.

%\bibliography{topo}

\begin{thebibliography}{99}
\bibitem{UnitsOfSigmaxy} Here $\sigma_{xy}$ is given in units of $({1\over 2\pi})(e^2/h)$.
\bibitem{FootnoteWZWlevelInteger} This ensures that two different extensions of ${\vec n}(\tau)$
yield that same contribution to the partition function.
\end{thebibliography}

%%%%%%%%%%%%%%%%%%%%%%%%%%%%%%%%%%%%%%%%%%%%%%%%%%%
% Acknowledgments
%%%%%%%%%%%%%%%%%%%%%%%%%%%%%%%%%%%%%%%%%%%%%%%%%%%

\paragraph{Acknowledgments.--}
This work has been supported in part, by NSF DMR-0706140
(A.W.W.L.). C. Xu is supported by the Alfred P. Sloan Foundation,
the David and Lucile Packard Foundation, Hellman Family
Foundation, and NSF Grant No. DMR-1151208. Both authors
acknowledge the hospitality of KITP as organizers of the programs
on `Topological Insulators and Superconductors' (A.W.W.L.) and
`Holographic Dualities (C.X.), and support from the National
Science Foundation at under Grant No. NSF PHY05-51164 (KITP).

%%%%%%%%%%%%%%%%%%%%%%%%%%%%%%%%%%%%%%%%%%%%%%%%%%%
% BIBLIOGRAPHY
%%%%%%%%%%%%%%%%%%%%%%%%%%%%%%%%%%%%%%%%%%%%%%%%%%%

\end{document}